\documentclass[12pt,fleqn,epsfig]{article}
\newcommand{\beq}{\begin{equation}}
\newcommand{\eeq}{\end{equation}}
\newcommand{\bqa}{\begin{eqnarray}}
\newcommand{\eqa}{\end{eqnarray}}

\usepackage{epsfig}
\usepackage{fleqn,espcrc1}
\usepackage{wrapfig}
\def\square{\vcenter{\vbox{\hrule height.4pt
          \hbox{\vrule width.4pt height8pt
          \kern8pt\vrule width.4pt}\hrule height.4pt}}}

\begin{document}
%\centerline{\Large\bf Thermal Field Theory in Equilibrium}
%\vskip 10mm
%\centerline{Jens O. Andersen}
%\centerline{\it Department of Physics, The Ohio State University,
%Columbus, OH 43210}
\vskip 3mm

\title{Thermal Field Theory in Equilibrium}
\author{Jens O. Andersen,\address{Physics Department, Ohio State University, Columbus OH 43210, USA}\thanks{Invited Talk given at 5th Workshop on QCD, 
Villefranche-sur-Mer, France 3-7 Jan 2000.}}

%\author{Jens O. Andersen\\ \it
%Physics Department, Ohio State University, Columbus OH 43210, USA}

\maketitle

\begin{abstract}
{\footnotesize }
In this talk, I review recent developments in equilibrium thermal
field theory. Screened perturbation theory and 
hard-thermal-loop perturbation theory are discussed.
A self-consistent $\Phi$-derivable approach is also briefly reviewed. 
\end{abstract}
%\small
%\newpage
\section{Introduction}
Thermal field theory has applications in many areas of physics
and one of those is the early Universe.
There is an excess of matter over antimatter in the present Universe. Unless
this was an initial condition in the Big Bang Scenario, this baryon asymmetry
must have been created during the evolution of the Universe.
According to Sakarov's three criteria for baryogenesis, the Universe must 
have been out of equilibrium, there must be baryon-number violating processes,
and there must be CP violation. 
Although baryon-number violating processes
are exponentially suppressed at $T=0$, 
they are significant at high temperatures. 
Moreover, CP violation is known from the kaon system, and 
the Universe was out of equilibrium during a
cosmological phase transition if it was first order. Hence, all the 
necessary ingredients of baryosynthesis may have been present in the
early Universe and has been subject of intense investigation in recent years.

Another important application of thermal field theory is heavy-ion collisions.
QCD is expected to undergo a phase transition at high temperature and/or
high density, where chiral symmetry is restored, and quarks and gluons
are deconfined. Hadrons are no longer the relevant degrees of freedom, but
matter is described in terms of a plasma of interacting
quarks and gluons. A quark-gluon
plasma is expected to be created in heavy-ion collisions at RHIC and LHC.
Signatures of the formation of a quark-gluon plasma includes photon and 
dilepton production, and $J/\psi$ suppression, 

In order to understand such complicated nonequilibrium phenomena, 
we need to have equilibrium field theory under control.
In this talk, I would like to give an overview of recent developments in 
equilibrium thermal field theory.
\section{Weak-coupling Expansion}
Let us start the discussion by considering a massless
scalar field theory
with a $\phi^4$ interaction. The Euclidean Lagrangian is
\bqa
{\cal L}%_{\mbox{\scriptsize $E$}}
&=&{1\over2}(\partial_{\mu}\Phi)^2+{g^2\over24}\Phi^4\;.
\eqa
Using ordinary perturbation theory, one splits ${\cal L}$ into a
a free part (quadratic piece) and treats the $\phi^4$ term as an interaction.
%\bqa
%{\cal L}_E^{\rm free}&=&{1\over2}(\partial_{\mu}\Phi)^2\;,\\
%{\cal L}_E^{\rm int}&=&{g^2\over24}\Phi^4\;.
%\eqa
The loop expansion is then an expansion in powers of $g^2$ around
an ideal massless gas. However,
it is well known that naive perturbation theory breaks down beyond two-loop 
order due to infrared divergences since the scalar field is massless,
and that one needs to reorganize the perturbative expansion. Physically, the infrared divergences are screened due to
a thermally generated mass of order $gT$. 
\begin{figure}[htb]
%\vspace*{-4.0cm}
\begin{center}
\hspace{1cm}
\epsfysize=0.7cm
%\begin{center}
\mbox{\psfig{figure=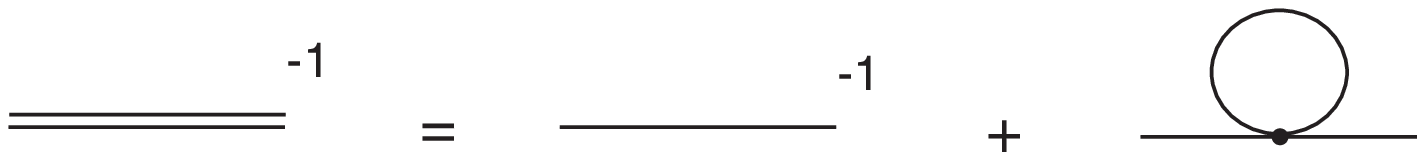,width=7cm,height=1cm}}
%\centerline{\epsffile{jensSD.eps}}
\vspace*{-0.4cm}
\caption[a]{Two-point function in the one-loop approximation.}
\label{}
\vspace*{-0.8cm}
\end{center}
\end{figure}

\begin{figure}[htb]
%\vspace*{-4.0cm}
\begin{center}
%\hspace{1cm}
\epsfysize=1cm
\epsffile{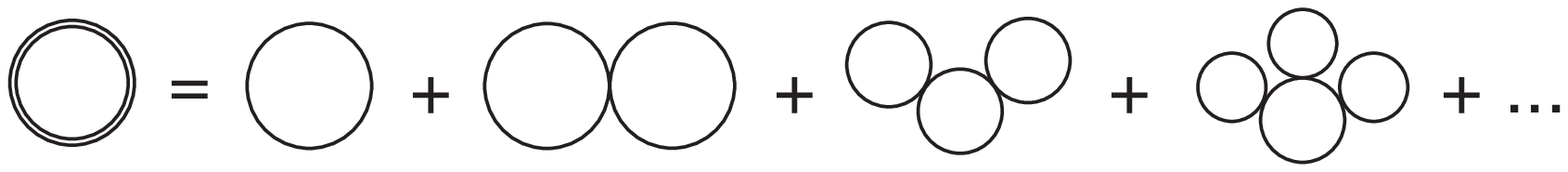}
\vspace*{-0.6cm}
\caption[a]{Summation of bubble diagrams using a dressed propagator.}
\label{bubble}
\end{center}
\vspace*{-0.8cm}
\end{figure}

In order to incorporate the physics of Debye 
screening, we need to use an effective
propagator that includes the mass. Using the effective propagator in a
one-loop calculation is equivalent to summing all the bubble diagrams of
Fig.~\ref{bubble}.

The summation of the bubble diagrams gives a contribution to the free energy
of order $g^3$, and is thus nonanalytic in $g^2$.
The resummation of diagrams can be made into a systematic expansion in powers
of $g$.
The free energy has been calculated through order $g^5$
in the weak-coupling expansion~\cite{arnold-zhai}:
\bqa
{\cal F}={\cal F}_{\rm ideal}\left[1-{5\over4}\left({g\over4\pi}\right)^2
+{5\sqrt{6}\over3}\left({g\over4\pi}\right)^3+{3\over2}\left({g\over4\pi}\right)^4
\right.
\left.+\left(24.5\log{g\over4\pi}+13.2\right)\left({g\over4\pi}\right)^5
%+...%{\cal O}(g^6\log g)
\right]\;,
\eqa
where ${\cal F}_{\rm ideal}=-(\pi^2/90)T^4$, and the renormalization scale
$\mu_4=2\pi T$. 
It turns out that the weak-coupling expansion does not
converge unless the coupling is tiny. 
This lack of convergence is not specific to $\phi^4$ theory but occurs also
in QCD~\cite{arnold-zhai}. 
In QCD, the $g^3$ term is smaller than the $g^2$ term
in the free energy only if $\alpha_s\sim 1/20$. This corresponds to a 
temperature of $10^5$ GeV, which
is many orders of magnitude larger than the temperatures expected in
heavy-ion collisions (approximately 0.5 GeV at RHIC).
\section{Screened Perturbation Theory}
There are several ways of reorganizing perturbation theory to improve its
convergence properties. One of the most successful approaches is 
``screened perturbation theory'' developed by 
Karsch,  Patk\'os, and  Petreczky~\cite{K-P-P}.
A local mass term is added to and subtracted from the Lagrangian, with the
added mass term treated nonperturbatively, and the subtracted term
as a perturbation. Thus the Lagrangian is split according to
\bqa
{\cal L}_{\rm free}&=&{1\over2}(\partial_{\mu}\phi)^2
+{1\over2}m^2\phi^2\;,\\
{\cal L}_{\rm int}&=&{g^2\over24}\phi^4-{1\over2}m^2\phi^2\;.
\eqa
Hence, screened perturbation theory is essentially expanding around
an ideal gas of massive particles. 
A straightforward calculation give the renormalized
one-loop free energy in screened
perturbation theory:
\begin{eqnarray} \nonumber
{\cal F_{\rm SPT}} &=&{1 \over 2}T 
\sum_{n = -\infty}^{\infty}\int{d^{3-2\epsilon}k\over(2\pi)^{3-2\epsilon}}\log\left( {\omega_n^2 + k^2 + m^2} \right)\\
&=&{T\over2\pi^2}\int_0^{\infty}dk\;k^2\log\left(1 -  e^{- \beta \omega} \right) +
{m^4\over32\pi^2}\left[{3\over4}+\log{\mu\over m}\right]\;,
\eqa
where $\omega_n=2\pi T$ are the Matsubara frequencies and $\mu$ is a
renormalizaton scale associated with dimensional regularization.

At this point I would like to emphasize that the screening mass $m$
is a completly arbitrary parameter. To complete a calculation using
screeened perturbation theory, one must specify how $m$ is determined.
Karsch {\it et al}.~\cite{K-P-P} used a one-loop gap equation
to determine the screening mass:
\bqa
\label{gap}
m^2(T)={g^2\over 4\pi^2}\int_{0}^{\infty}dk\;
{k\over e^{\beta \omega}-1}\;,\hspace{1cm}
{\rm where}\;\;\omega=\sqrt{k^2+m^2}\;.
\eqa

\begin{wrapfigure}[]{l}[0pt]{9cm}
%\begin{figure}[htb]
\vspace*{-1.4cm}
\begin{center}
%\hspace{1cm}
\epsfysize=6.1cm
\epsffile{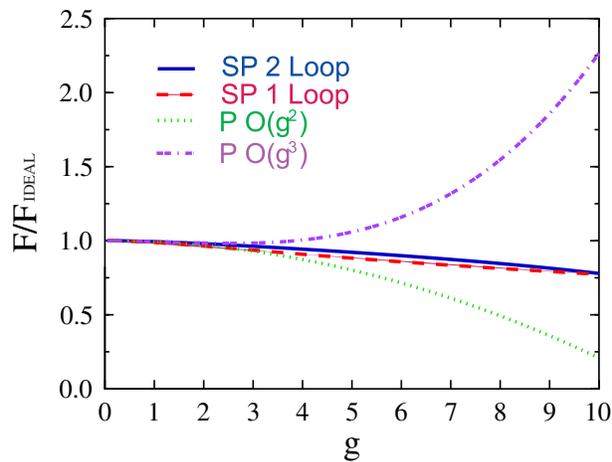}
\caption[a]{Comparison of screened perturbation theory (SP) and the 
weak-coupling expansion (P).}
%\vspace*{-0.3acm}
\label{screfig}
\end{center}
%\end{figure}
\end{wrapfigure}

In Fig.~\ref{screfig}, we show the weak-coupling expansions through
order $g^2$ (lower curve) and order $g^3$ (upper curve)
normalized to ${\cal F}_{\rm ideal}$.
The two approximations to the free energy have different signs
and show the lack of convergence of the weak-coupling expansion.
The two curves that are almost on top of each other are the one and two-loop
approximations in screened perturbation theory, with the mass parameter
determined from Eq.~(\ref{gap}). We conclude that screened 
perturbation theory has good convergence properties
for a wide range of values for the coupling constant $g$.
%\newpage
\section*{\hspace{-9.32cm}4. HTL Perturbation Theory}
\hspace{-9.32cm}
We would like to generalize screened perturbation theory to gauge theories.
We cannot simply add and subtract a local mass term in the Lagrangian
since this would violate
gauge invariance. However, there is a way to incorporate plasma effects,
including propagation of massive quasiparticles, screening of interactions
and Landau damping and still maintain gauge invariance. This approach
is hard-thermal-loop (HTL) perturbation theory, and involves effective 
propagators and effective vertices~\cite{bp,EJM1}.

The free energy of pure-glue QCD to leading order in HTL perturbation theory is
\bqa
{\cal F}= 8\left[(d-1) {\cal F}_T + {\cal F}_L+ \Delta{\cal F} \right]
\, ,
\eqa
where $d=3-2\epsilon$, ${\cal F}_T$ and ${\cal F}_L$
are the transverse and longitudinal contributions to the free energy, 
respectively,
and $\Delta{\cal F}$ is a counterterm.
In the imaginary-time formalism, we have
\begin{eqnarray}
{\cal F}_T &=& {1 \over 2 \beta} 
%\mu^{3-d}
\sum_n \int_k \log\left[k^2 + \omega_n^2+\Pi_T(\omega_n,k)\right]\;,\hspace{1cm}
\\
{\cal F}_L &=& {1 \over 2 \beta} 
%\mu{3-d}
\sum_n \int_k \log\left[k^2 - \Pi_L(\omega_n,k)\right] \, ,
\end{eqnarray}
where the transverse and longitudinal self-energy functions are
\bqa
\Pi_T&=& -{3 \over 2} m_g^2 {\omega_n^2 \over k^2} \left[ 1 - 
			{\omega^2_n +k^2 \over 2 i\omega_n k} \log {i\omega_n + k \over i\omega_n - k} \right]\;,%\hspace{0.35cm} 
\\
\Pi_L&=& 3 m_g^2 \left[ {i\omega_n \over 2 k} \log {i\omega_n + k \over i\omega_n - k} - 1 \right]\;.
\eqa
The sum over the Matsubara frequencies $\omega_n=2\pi nT$ can be rewritten 
as a contour integral
around a contour $C$ that encloses the points $\omega=i\omega_n$. 
The integrand has branch cuts that start at $\pm\omega_T(k)$ and
$\pm\omega_L(k)$, where $\omega_T(k)$ and $\omega_L(k)$
are the dispersion relations for transverse and longitudinal 
gluon quasiparticles, respectively. The integrand also
has a branch cut running from $\omega=-k$ to $\omega=k$ due to the functions
$\Pi_T$ and $\Pi_L$.
The contour
can be deformed to wrap around the quasiparticle and Landau-damping
branch cuts. Some of the temperature-independent integrals over $\omega$ 
can be
calculated analytically, while others must be evaluated numerically. 
With dimensional regularization, the logarithmic ultraviolet 
divergences show up as poles in $\epsilon$.
Using the modified minimal subtraction ($\overline{\mbox{MS}}$)
renormalization prescription
with the counterterm $\Delta{\cal F}=9m^4_g/64\pi^2\epsilon$, we obtain

\begin{eqnarray} \nonumber
{\cal F}_{\rm HTL} &=&{8\over \pi^2}T \int^\infty_0 k^2 dk \log (1-e^{-\beta \omega_T})  + {4 \over \pi^2}T\int^\infty_0 k^2 dk \log {1-e^{-\beta \omega_L} \over 1-e^{-\beta k}}+{1\over2}\; m_g^2 T^2 \\
%		\nonumber \\
	&&-{8 \over \pi^3} \int^\infty_0 d\omega\;
%		{1 \over e^{\beta \omega} -1} 
n(\omega)
\int^\infty_\omega k^2 dk 
		\left [ 2\phi_T - \phi_L\right] 
%+  {1 \over 2} 
+ {9 \over8 \pi^2}\; m_g^4 \; 
		\left[\log {m_g \over \mu_3} + 0.31 \right] \, .
\label{htl}
\end{eqnarray}
Here, $\omega_T$ and $\omega_L$ are the transverse and longitudinal
dispersion relations which are the solutions to
$k^2-\omega_T^2+\Pi_T(-i\omega_T,k)=0$, and
$k^2-\Pi_L(-i\omega_L,k)=0$.
Moreover, $n(\omega)$ is the Bose-Einstein distribution 
function and 
the angles $\phi_T$ and $\phi_L$ satisfy
\bqa
{3 \pi \over 4} m^2_g {\omega (k^2-\omega^2) \over k^3} 
	\cot \phi_T & = & 
k^2-\omega^2 \;+\; {3 \over 2} m^2_g {\omega^2 \over k^2}
\left[ 1 + {(k^2-\omega^2) \over 2 k \omega}\log{k+\omega\over k-\omega}
\right]\;,
\label{theta-T} 
\\ 
{3 \pi \over 2} m_g^2 {\omega \over k} \cot \phi_L &=& 
k^2 \;+\; 3m_g^2 
\left [ 1 - {\omega \over 2 k} \log{k+\omega\over k-\omega} \right]\;.
\label {theta-L}
\eqa

\newpage
\begin{wrapfigure}[]{l}[0pt]{8cm}
%\begin{figure}[htb]
%\vspace*{-1.2cm}
%\begin{center}
%\hspace{1cm}
\epsfysize=6.1cm
\epsffile{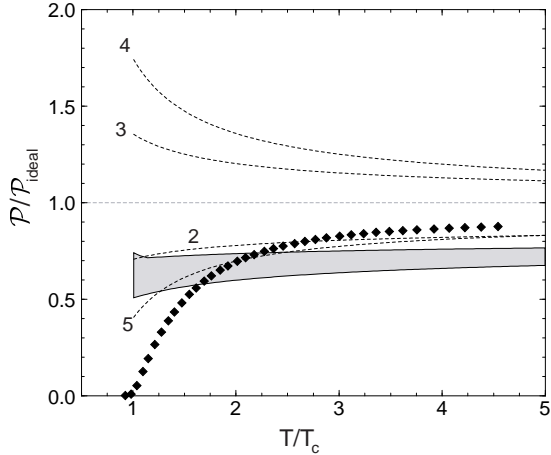}
\caption[a]{HTL pressure for pure-glue QCD as a function of
$T/T_c$. Lattice results from Ref.~\cite{boyd}.}
\label{pres}
\vspace*{-0.8cm}
%\end{center}
\end{wrapfigure}

The leading-order HTL result for the pressure is shown in 
Fig.~\ref{pres} as the shaded band that corresponds to varying the
renormalization scales $\mu_3$ and $\mu_4$ 
by a factor of two around their central values
$\mu_3=0.717m_g$ and $\mu_4=2\pi T$.
This value of $\mu_3$ is chosen in order to minimize the 
pathological behavior of ${\cal F}_{HTL}$ at low temperatures~\cite{EJM1}.
We also show as dashed curves the weak-coupling expansions through order
$\alpha_s$, $\alpha_s^{3/2}$, $\alpha_s^2$, and $\alpha_s^{5/2}$ labelled
2, 3, 4, and 5.
We have used a parameterization of the running coupling constant 
$\alpha_s(\mu_4)$ that includes
the effects of two-loop running
%, and the renormalization scale 
%is $\mu_4=2\pi T$. 
%The renormalization scale
%$\mu_3$ arises because the HTL free energy has logarithmic ultraviolet
%divergences, and its origin is three-dimensional.
With the above choices of the renormalization scales, 
our leading-order result for the HTL free energy lies 
below the lattice results of Boyd {\it et al}.~\cite{boyd} 
(shown as diamonds) for $T>2T_c$.
However, the deviation from lattice QCD results
has the correct sign and roughly the correct magnitude to be
accounted for by next-to-leading order corrections in HTL perturbation 
theory~\cite{EJM1}.
Comparing the weak-coupling expansion with the 
the high-temperature expansion of~(\ref{htl}),
and identifying $m_g^2$ with its weak-coupling limit 
${4\pi\over3}\alpha_sT^2$, 
we conclude
that HTL perturbation theory overincludes the $\alpha_s$ contribution
by a factor of three~\cite{EJM1}. The $\alpha_s^{3/2}$ contribution 
which is associated with Debye screening is included correctly. 
At next-to-leading order, HTL perturbation theory agrees with the weak-coupling
expansion through order $\alpha_s^{3/2}$. 
Thus the next-to-leading order contribution to
${\cal F}_{}/{\cal F}_{\rm ideal}$ in HTL perturbation theory will be positive
at large $T$ since it must approach $+{15\over2}\alpha_s/\pi$.

\section{Self-consistent $\Phi$-derivable Approach}
%\vspace{-1cm}
An alternative to screened perturbation theory is 
the self-consistent $\Phi$-derivable approach~\cite{phid}. 
The free energy can be expressed as the stationary point of the 
thermodynamic potential $\Omega[D]$.
\bqa
\beta\Omega[D]&=&{1\over2}{\rm Tr}\log D^{-1}-{1\over2}\Pi D+\Phi[D]\;,
\eqa
where $D$ is the exact propagator and
$\Phi[D]$ is given by the sum of two-particle irreducible diagrams.
the condition that 
$\Omega[D]$ be a statinary point
gives an integral equation for the propagator:
\bqa
{\partial\Phi[D]\over\partial D}={1\over2}\Pi\;.
\eqa
The free energy ${\cal F}$ is obtained by
solving the integral equation for $D$
and inserting the solution into the the thermodynamics potential $\Omega[D]$. 
Blaizot {\it et al}.~\cite{gauge} have 
developed a HTL approximation to the 
$\Phi$-derivable approach and applied it to both scalar theory and 
nonabelian gauge theories.

I do not have enough time to discuss this approach in depth
and compare it critically with 
screened perturbation theory and HTL perturbation theory, but I will mention
a few key points:
\begin{itemize}
\item{The $\Phi$-derivable approach is 
thermodynamically consistent, which
means that the usual thermodynamic relations between pressure, energy density
and entropy hold exactly. Thermodynamic consistency is destroyed by the
HTL approximation of Ref.~\cite{gauge}.
It holds only up to perturbative corrections in screened perturbation
theory and HTL perturbation theory.}
\item{The $\Phi$-derivable free energy
is gauge dependent since only the propagator and not the vertices are
dressed. This problem is avoided in the HTL approximation of Tef~\cite{gauge}
by not solving the gap equation with sufficient accuracy the see the 
gauge dependence.
HTL perturbation theory is gauge-fixing independent by construction.}
\item{The running of the coupling constant in the $\Phi$-derivable approach
is not consistent with the $\beta$ function of the theory.
In Ref~\cite{gauge}, the correct running is put in by hand.}
\end{itemize}
Higher-order calculations using the self-consistent $\Phi$-derivable
approach are currently being 
carried out for the scalar theory by Braaten and Petitgirard~\cite{petit}.
\section{Summary}
The weak-coupling expansion is useless for temperatures which are relevant
for experiments at RHIC and LHC. 
We have seen that screened perturbation theory shows good convergence for
a large range of values for the coupling $g$.
This fact gives us hope that HTL perturbation theory
might be a useful framework for calculating static and dynamical 
quantities of a quark-gluon plasma at experimentally accessible energies.
A next-to-leading order calculation of the free energy
using HTL perturbation theory
is currently being carried out~\cite{two}.
\section*{Acknowledgments}
The work on HTL perturbation theory has been done in collaboration with 
Eric Braaten and Michael Strickland~\cite{EJM1}.
The author would like to thank the organizers of 5th workshop on QCD for
an interesting and stimulating meeting. 
This work was supported in part by a Faculty Development Grant 
from the Physics Department of the Ohio State University.
%and by a fellowship from the Norwegian Research Council 
%(project 124282/410).


\begin{thebibliography}{99}
%\bibitem{schiff}D. Schiff, these proceedings.
\bibitem{arnold-zhai}
P. Arnold and C. Zhai, Phys. Rev. {\bf D50}, 7603 (1994);
Phys. Rev. {\bf D51}, 1906 (1995).
\bibitem{K-P-P}
F. Karsch, A. Patk\'os, and P. Petreczky, Phys. Lett. {\bf B401}, 69 (1997).
\bibitem{bp}E. Braaten and R.D. Pisarski,  Phys. Rev. Lett. {\bf 64}, 1338 (1990);
	Nucl. Phys. {\bf B337}, 569 (1990).
\bibitem{EJM1} J. O. Andersen, E. Braaten and M. Strickland, Phys. Rev. Lett. {\bf 83}, 2139 (1999); Phys Rev. {\bf D61}, 014017 (1999); hep-ph/9908323.
\bibitem{boyd} G. Boyd et al., Phys. Rev. Lett. {\bf 75} 4169 (1995);
        Nucl. Phys. {\bf B469} 419 (1996).
\bibitem{phid}J. M. Luttinger and J. C. Ward, {\bf 118}, 1417 (1960);
G. Baym, Phys. Rev. {\bf 127}, 1391 (1962). 
\bibitem{gauge}J.-P. Blaizot, E. Iancu and A. Rebhan, Phys. Rev. Lett. {\bf 83} (1999),  hep-ph/9910309. 
\bibitem{petit} E. Braaten and E. Petitgirard, in preparation.
\bibitem{two}J. O. Andersen, E. Braaten and M. Strickland, 
in preparation.
\end{thebibliography}
\end{document}